\documentclass[preprint,11pt,showkeys,nofootinbib,floatfix]{revtex4}
\usepackage{amssymb}
\usepackage{multirow}
\usepackage{amsfonts}
\usepackage{amsmath}
\usepackage{graphicx}
\usepackage{subfigure}
\usepackage[dvipdfm,  
            pdfstartview=FitH,
            bookmarksnumbered=true,
            bookmarksopen=true,
            colorlinks,
            pdfborder=001,
            linkcolor=green,
            anchorcolor=green,
            citecolor=red
            ]{hyperref}
\usepackage{graphicx}
\usepackage[toc,page,title,titletoc,header]{appendix}

\begin{document}
\title{Lifshitz effects on holographic $p$-wave superfluid}
\author{Ya-Bo Wu$^{1}$}
\thanks{E-mail address:ybwu61@163.com}
\author{Jun-Wang Lu$^{1}$}
\author{Cheng-Yuan Zhang$^{1}$}
\author{Nan Zhang$^{1}$}
\author{Xue Zhang$^{1}$}
\author{Zhuo-Qun Yang$^{1}$}
\author{Si-Yu Wu$^{1}$}
\affiliation{
$^{1}$Department of Physics, Liaoning Normal University, Dalian, 116029, China}
\begin{abstract}
In the probe limit, we numerically build a holographic $p$-wave superfluid model in the four-dimensional Lifshitz black hole coupled to a Maxwell-complex vector field.
We observe the rich phase structure and find that the Lifshitz dynamical exponent $z$ contributes evidently to the effective mass of the matter field and dimension of the gravitational background.  Concretely, we obtain the Cave of Winds  appeared only in the five-dimensional anti-de Sitter~(AdS) spacetime,  and the increasing $z$ hinders not only the condensate but also the appearance of the first-order phase transition. Furthermore,  our results agree with the Ginzburg-Landau results near the critical temperature. In addition, the previous AdS superfluid model is generalized to the Lifshitz spacetime.
\end{abstract}
\maketitle
\section{Introduction}
 The gauge/gravity duality~\cite{Maldacena1998} builds a powerful relationship between the many-body system in the quantum mechanics with the strong interaction and the classical dynamical black hole with a higher dimension of spacetime, hence, in the past decades, it has been widely used to study various strongly coupled system. In particular, via this holographic duality, various high-temperature superconductors\footnote{According to the gauge/gravity duality, there is not dynamical gauge field in the dual field theory~\cite{Hartnoll:2008kx}. Therefore, the current induced by the applied magnetic field  can not produce an equal and opposite canceling field in the superconductor to exclude the external magnetic field, which is different from the ordinary superconductor but rather similar to thin superconducting films or wires.} were constructed, which involve different gravitational  backgrounds as well as matter fields, see, for example, Refs.~\cite{Hartnoll2008,Gubser2008a,Chen:2010mk,Hartnoll:2008kx,Cai:2014oca,Zeng:2014uoa,Cai:2014jta,Albash:2008eh,Liu:2011fy,Nishioka131,
 Momeni:2012ab,Roychowdhury:2012vj,Herzog:2009xv,Horowitz:2010gk} and the references therein.

All above superconductor models almost base on the isotropic gravitational backgrounds. Due to the various anisotropy of superconductors in the condensed matter system, the authors of Ref.~\cite{Kachru} proposed a $(d+2)$-dimensional gravity  dual to the  Lifshitz anisotropic scaling of the space and time, $ds^2=L^2\left(-r^{2z}dt^2+r^2d\vec{x}^2+\frac{dr^2}{r^2}\right)$,
where $d\vec{x}^2=dx^2_1+\cdots+dx^2_d$, $r\in(0,\infty)$ and $z$ the dynamical critical exponent as well as $L$ a cosmological constant. In particular, the Lifshitz fixed points scale the space and time as
$t\rightarrow b^z t, \vec{x}\rightarrow b \vec{x}$ ($z\neq 1$). For related works, see also, for instance, Refs.~\cite{Horava:2009uw,Liu:2013kla,Liu:2014lga}. In addition, the gravity duality with the anisotropy between two spatial directions also exists, see, for example, Refs.~\cite{Iizuka:2012wt,Koga:2014hwa}. In the remainder of this paper, we will set $L=1$ for simplicity.  Subsequently, the Lifshitz spacetime was extended to a $(d+2)$-dimensional  finite-temperature system~\cite{Pdw09052678}
\begin{equation}\label{Lifm}
ds^2=-r^{2z}f(r)dt^2+\frac{dr^2}{r^2f(r)}+r^2\sum^{d}_{i=1}dx_i^2, \qquad f(r)=1-\frac{r_0^{z+d}}{r^{z+d}},
\end{equation}
where $r_0$ denotes the location of the event horizon. Moreover, the Hawking temperature can be written as $T=\frac{(z+d) r_0^z}{4\pi}$. To see the anisotropic effects, some holographic superconductors were constructed in the Lifshitz black hole backgrounds, see, for example, Refs.~\cite{Brynjolfsson065401,Sin4617,Buyanyan,Lu:2013tza,Zhao:2013pva,Momeni:2012tw,Lala:2014jca,
Tallarita:2014bga,Lin:2014bya,Guo:2014wca,Jing:2014bza}, where the results showed that the larger Lifshitz parameter $z$ hinders the condensate. What is more, the Lifshitz parameter $z$ contributes to the effective dimension of the gravitational background.

In order to generalize the above superconductor models to the ones with a steady current, holographic superfluid solutions were constructed by performing a deformation to the superconducting black hole~\cite{Basu:2008st,Herzog:2008he}, and were further investigated in Refs.~\cite{Amado:2009ts,Arean:2010xd,Arean:2010zw,Zeng:2010fs, Sonner:2010yx,Amado:2013xya,Amado:2013aea}. It follows that below the critical temperature $T_0$ with the vanishing superfluid velocity, there is a special value of $T$, beyond (below) which the order of the phase transition is of second (first) order. We call the critical superfluid velocity corresponding to this special temperature as the translating superfluid velocity. Moreover, in the five-dimensional~(5D) anti-de Sitter~(AdS) black hole background, the authors of Ref.~\cite{Arean:2010zw} found that  when the temperature decreases, the second-order transition occurs before the first-order transition to a new superconducting phase.

Recently, motivated by Ref.~\cite{Hartnoll2008}, a new holographic $p$-wave superconductor model was built by coupling a Maxwell-complex vector (MCV) field with the four-dimensional~(4D) Schwarzschild AdS black hole~\cite{Cai:2013pda}, for related works, see Refs.~\cite{Cai:2013aca,Li:2013rhw,Cai:2014ija, Cai:2013kaa,Wu:2014dta,Wu:2014lta,ybwu2014}. It was shown that only the external magnetic field can induce the condensate, which is similar to result of the QCD vacuum phase transition in Ref.~\cite{Chernodub:2011mc} compared with ones in Ref.~\cite{Bali:2011qj}. In addition, even in the Lifshitz spacetime, this MCV model is still a generalization of the usual $p$-wave model realized by the SU(2) Yang-Mills (YM) gauge field~\cite{Gubser2008a}. Because of the anisotropic properties of the superfluid model in the real world, for example, the $He_3$ superfluid, it is valuable to construct the holographic $p$-wave superfluid model by coupling the MCV model in the 4D Lifshitz black hole. More interesting questions are whether we can see (i) the Cave of Winds only existed in the 5D AdS black hole when considering the effects of Lifshitz parameter $z$ on the dimension of the gravitational background; (ii) the disappearance of the first-order phase transition due to the fact that the larger parameter $z$ hinders the condensate. Answering these questions is just the purpose of this paper.

Based on the above mentioned, we will build a holographic superfluid model in the 4D Lifshitz black hole coupled with the MCV field in the probe limit. Interestingly, we obtain the rich structure, especially the Cave of Winds, which means that the Lifshitz parameter $z$ contributes evidently to the effective mass of the matter field and the dimension of the background spacetime. Moreover, the larger $z$ not only decreases the critical temperature, but also hinders the emergence of the first-order phase transition.

The paper is organized as follows. In Sec.~II, we obtain the equations of motion and the grand potential for the superfluid model.  We numerically study the condensate and the supercurrent in Sec.~III and~IV, respectively. The last section is devoted to  the conclusions and further discussions.

\section{Equations of motion and the grand potential }
In this section, we derive the equations of motion in terms of the MCV field, following which we obtain the grand potential.

The MCV matter action  including a Maxwell field and a complex vector field reads~\cite{Cai:2013aca}
\begin{equation}\label{Lvector}
\mathcal{S}_m=\frac{1}{16\pi G_{4}}\int dx^{4}\sqrt{-g}\left(-\frac{1}{4}F_{\mu\nu}F^{\mu\nu}-\frac{1}{2}\rho_{\mu\nu}^\dag\rho^{\mu\nu}-
m^2\rho^\dag_\mu\rho^\mu+iq \gamma \rho_\mu\rho^\dag_\nu F^{\mu\nu}\right),
\end{equation}
where $F_{\mu\nu}$ is the strength of the Maxwell field $A_\mu$ and $\rho_{\mu\nu}=D_\mu\rho_\nu-D_\nu\rho_\mu$ with the covariant derivative $D_\mu=\nabla_\mu-iq A_\mu$, while $m$  and $q$  are the mass  and the charge of the vector field $\rho_\mu$, respectively.
The last term with a coefficient $\gamma$ stands for the  interaction between $\rho_\mu$ and $A_\mu$, which is crucial to the effect of the magnetic field in the holographic model~\cite{Cai:2013pda,Cai:2013kaa,Wu:2014dta,Wu:2014lta}. However, in this paper we do not consider  the magnetic field, hence, it will not contribute to our work. Moreover, we will work in the probe limit that can be realized by taking $q\rightarrow\infty$ with $q\rho_\mu$ and $qA_\mu$ fixed.

By varying the action (\ref{Lvector}), we obtain the equations of motion
\begin{eqnarray}
 D^\nu\rho_{\nu\mu}-m^2\rho_\mu+iq\gamma\rho^\nu F_{\nu\mu}&=&0,\label{EOMrho}\\
 \nabla^\nu F_{\nu\mu}-iq(\rho^\nu\rho^\dag_{\nu\mu}-\rho^{\nu\dag}\rho_{\nu\mu})
  +iq\gamma\nabla^\nu(\rho_\nu\rho^\dag_\mu -\rho^\dag_\nu\rho_\mu)&=&0.\label{EOMphi}
 \end{eqnarray}

As Ref.~\cite{ybwu2014}, we turn on the following ansatzs for  $\rho_\mu$ and  $A_\mu$
\begin{equation}\label{rhoA}
\rho_\nu dx^\nu=\rho_x(r) dx,\ \ \ \ \  A_\nu dx^\nu=\phi(r) dt+A_y(r)dy.
\end{equation}
Thus  the concrete equations of motion in terms of the matter field are given by
\begin{eqnarray}
\rho_x ''+\left(\frac{z+1}{r}+\frac{f'}{f}\right)\rho_x' -\frac{\rho_x}{r^2 f}\left(m^2-\frac{\phi ^2}{r^{2z} f} +\frac{A_y^2}{r^2}\right) &=&0, \label{Eqpsi}\\
\phi ''+\frac{3-z}{r}\phi '-\frac{2 \rho_x ^2 }{ r^4 f}\phi&=&0,\label{Eqphi}\\
A_y''+\left(\frac{z+1}{r}+\frac{ f'}{f}\right)A_y'-\frac{2 \rho_x ^2}{ r^4 f}A_y&=&0.\label{EqAx}
 \end{eqnarray}
When we turn off the spatial component $A_y(r)$, Eqs.~(\ref{Eqpsi}) and (\ref{Eqphi}) reduce to the ones in Ref.~\cite{Wu:2014dta}, while Eqs.~(\ref{Eqpsi}), (\ref{Eqphi}) and (\ref{EqAx}) with $z=1$ are the same with the ones in Ref.~\cite{ybwu2014}.

Due to the difficulty to solve the above equations analytically, here we turn to  the numerical approach, i.e., the shooting method~\cite{Herzog:2008he,Basu:2008st,Sonner:2010yx,Arean:2010xd,Zeng:2010fs,Arean:2010zw}. Before the numerical calculation, we should impose some boundary conditions on Eqs.~(\ref{Eqpsi}), (\ref{Eqphi}) and (\ref{EqAx}). In particular, at the horizon, $\rho_x(r_0)$ and  $A_y(r_0)$ are required to be regular, while $A_t(r_0)$ vanishing in order for the normal form of $g^{\mu\nu}A_\mu A_\nu$. At the infinity boundary $r\rightarrow\infty$, the general falloffs of the fields are of the forms
\begin{equation}\label{asyrpa}
\rho_x(r)=\frac{\rho_{x-}}{r^{\Delta_-}}+\frac{\rho_{x+}}{r^{\Delta_+}}+\cdots,\qquad
\phi(r)=\mu-\frac{\rho}{r^{2-z}}+\cdots,\qquad
A_y(r)=S_y-\frac{J_y}{r^z}+\cdots
\end{equation}
with $\Delta_\pm=\frac{1}{2}\left(z\pm\sqrt{z^2+4m^2}\right)$. According to the gauge/gravity duality,  $\rho_{x-}$ and $\rho_{x+}$ are usually interpreted as the source  and the vacuum-expectation value of the boundary operator $O_x$, respectively, while $\mu$, $\rho$, $S_y$, and $J_y$ as the chemical potential, the charge density, the superfluid velocity, and the supercurrent, respectively. To satisfy the requirement that the symmetry is broken spontaneously, we impose the source-free condition, i.e., $\rho_{x-}=0$.

There is a scaling symmetry for the asymptotical solutions~(\ref{asyrpa}) as $
 (r,S_y)\rightarrow \lambda (r,S_y),~(T,\mu)\rightarrow \lambda^z (T,\mu),~\rho_{x+}\rightarrow \lambda^{\Delta_++1} \rho_{x+},~J_y\rightarrow \lambda^{z+1} J_y$ and $\rho\rightarrow \lambda^2 \rho$ with $\lambda$ a positive real constant, by using which  we can fix the chemical potential and thus work in the grand canonical ensemble. As we know from  Refs.~\cite{Herzog:2008he,Basu:2008st,Arean:2010zw}, when the critical superfluid velocity increases beyond a translating value, the second-order phase transition will switch to the first-order one in the grand canonical ensemble. To determine which phase is more thermodynamically favored in this case, we should calculate the grand potential $\Omega$ of the bound state, which is identified with the Hawking temperature times the Euclidean on-shell action. From the action (\ref{Lvector}), the on-shell action $\mathcal{S}_{os}$ reads
\begin{eqnarray}
\mathcal{S}_{os}&=&\int dxdydtdr\sqrt{-g}\left(-\frac{1}{2}\nabla_\mu (A_{\nu}F^{\mu\nu})-\nabla_\mu(\rho^\dag_{\nu}\rho^{\mu\nu})+\frac{1}{2}A_\nu\nabla_\mu F^{\mu\nu}\right) \nonumber\\
&=&\frac{V_{2}}{T}\left(-\sqrt{-\gamma}n_r\rho^\dag_{\nu}\rho^{r\nu}|_{r\rightarrow\infty}-\frac{1}{2}
\sqrt{-\gamma}n_r A_\nu F^{r\nu}|_{r\rightarrow\infty}+\frac{1}{2}\int_{r_0}^\infty dr \sqrt{-g}A_\nu\nabla_\mu F^{\mu\nu}\right) \nonumber\\
&=& \frac{V_{2}}{T}\left(\frac{1}{2}((2-z)\mu\rho-z S_y J_y)+\int_{r_0}^\infty dr \psi^2\left(\frac{A_y^2}{r^{3-z}}-\frac{\phi^2}{r^{z+1}f}\right)\right),
\end{eqnarray}
where we have plugged the general falloffs (\ref{asyrpa}), and considered the integration $\int dt dxdy=\frac{V_2}{T}$ as well as  ignored the prefactor $\frac{1}{16\pi G_{4}}$ for simplicity. Since we work in the probe limit and impose the source-free boundary condition, we do not need to introduce the Gibbons-Hawking boundary term for the well-defined Dirichlet variational problem and the counterterms for the divergent terms in the on-shell action. For the mathematical simplicity, we usually work in the new coordinate $u=\frac{r_0}{r}$, therefore, the grand potentials in the superconducting phase $\Omega_{S}$ and the normal phase $\Omega_{N}$ are respectively~\cite{Basu:2008st,Zeng:2010fs}
\begin{eqnarray}
\frac{\Omega_{S}}{V_{2}}&=&\frac{1}{2}((z-2)\mu\rho+z S_y J_y)+\int_{\epsilon}^1 du \rho_x^2\left(\frac{u^{z-1}\phi^2}{1-u^{z+2}}-u^{1-z}A_y^2\right),\quad\frac{\Omega_N}{V_2}=-\frac{1}{2}\mu^2,\label{FreeE}
 \end{eqnarray}
 where the lower bound $u\rightarrow\epsilon$ corresponds to the boundary $r\rightarrow\infty$. Obviously, in the case of $z=1$, the grand potentials (\ref{FreeE}) reduce to the ones in Ref.~\cite{ybwu2014}.
\section{Condensates versus the temperature}
In this section, we calculate the condensate for different values of the Lifshitz parameter $z$ and the superfluid velocity with the fixed $\Delta_+=\frac{3}{2}$ and $2$, respectively.

\begin{figure}
\includegraphics[width=2.9in]{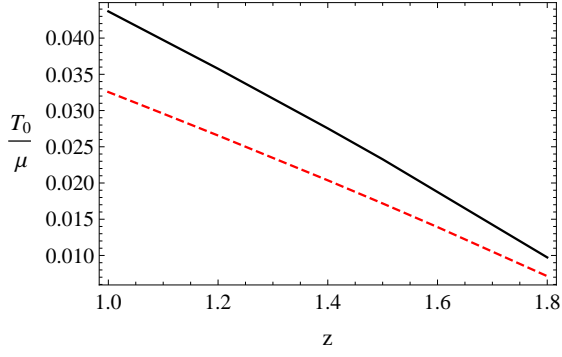}
\caption{The critical temperature versus the Lifshitz dynamical exponent $z$ in the absence of the superfluid velocity with  $\Delta_+=\frac{3}{2}$~(solid) and $2$~(dashed).}
\label{CriTz}
\end{figure}

As we know, in the absence of the superfluid velocity~\cite{Wu:2014dta}, the holographic conductor/superconductor phase transition is always the second-order one, hence, we can plot the critical temperature $T_0$ as a function of $z$ with  $\Delta_+=\frac{3}{2}$ and $2$ in Fig.~\ref{CriTz}.
It follows that for the fixed $\Delta_+$ ($z$), the critical temperature decreases with the increasing $z$~($\Delta_+$). In particular, in the case of $\Delta_+=\frac{3}{2}$, the results are consistent with the ones in Ref.~\cite{Wu:2014dta}, which means that the improving $z$ hinders the superconductor phase transition. This can be understood from the effective mass in Eq.~(\ref{Eqpsi}), i.e., $m^2_{eff}=m^2-\frac{\phi ^2}{r^{2z} f}$ that the increasing $z$ improves $m^2_{eff}$ so that we should further decrease the temperature to trigger the instability of the gravitational system.

\begin{figure}
\begin{minipage}[!htb]{0.45\linewidth}
\centering
\includegraphics[width=2.9in]{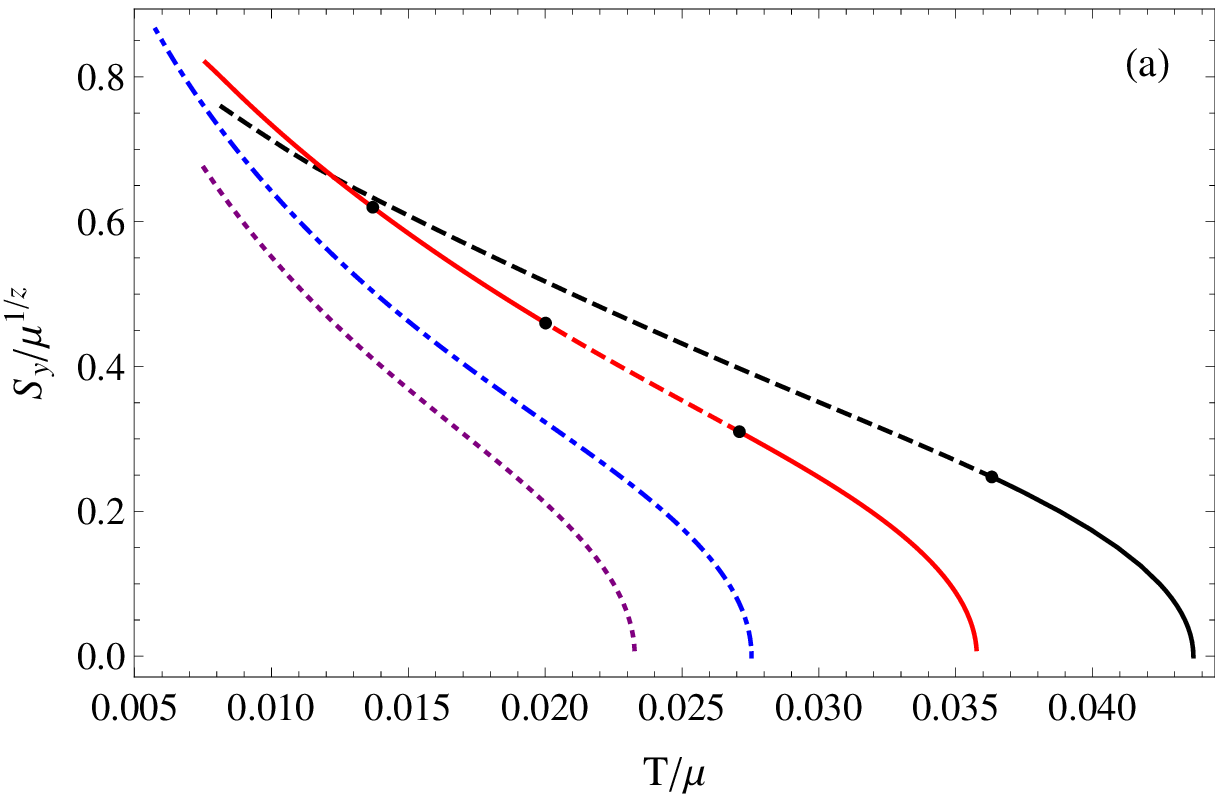}
\end{minipage}
\begin{minipage}[!htb]{0.45\linewidth}
\centering
\includegraphics[width=2.9in]{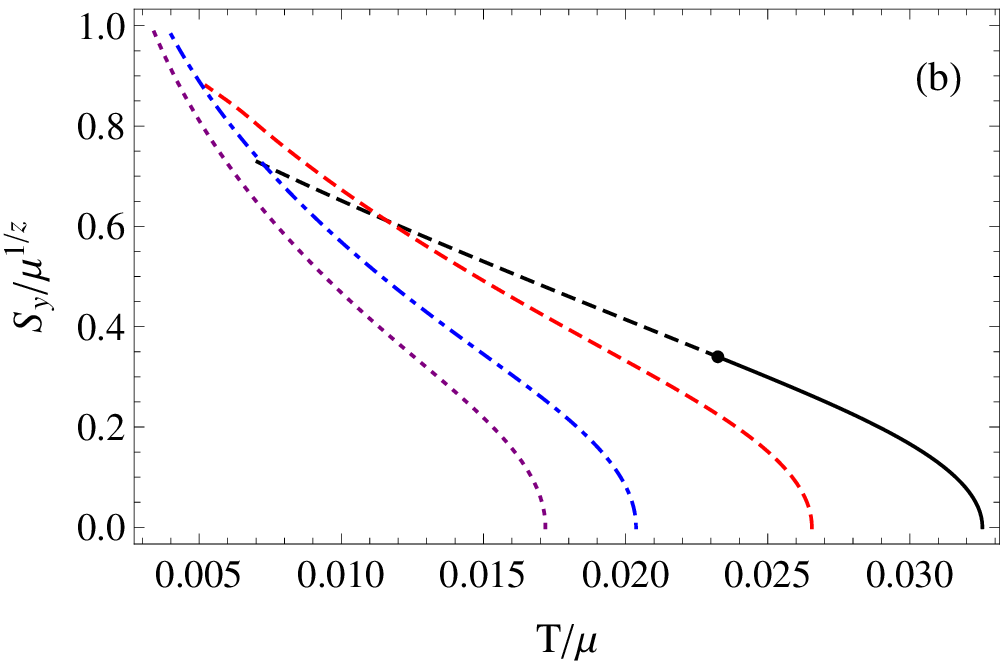}
\end{minipage}
\caption{The critical superfluid velocity versus the temperature with  $\Delta_+=\frac{3}{2}$~(a) and $2$~(b). The curves from top to bottom correspond to $z=1,~\frac{6}{5},~\frac{7}{5}$, and $\frac{3}{2}$, respectively.}
\label{TraPz}
\end{figure}

Taking into account the superfluid velocity, we can obtain the phase diagrams\footnote{It should be stressed that our phase diagrams are just from the state where the vector field begins to condensate but not consider whether the phase is thermodynamical favored or not. The following condensate will further complement the phase diagram. } with the different values of $z$ in Fig.~\ref{TraPz},
from which we have the following comments. The dependence of the critical value of $\frac{S_y}{\mu^{1/z}}$ on $\frac{T}{\mu}$ with $\Delta_+=2$ is generally similar to the one with $\Delta_+=\frac{3}{2}$. For the fixed temperature, the critical superfluid velocity decreases with the improving $z$, which implies that the larger the Lifshitz parameter $z$, the more easily the superfluid phase is broken into the normal phase. Conversely, the larger $z$ makes the superfluid phase transition more difficult. For the fixed superfluid velocity, the critical temperature decreases with the increasing $z$. Thus, we conclude that the larger $z$ hinders the superfluid phase transition whether the superfluid velocity is vanishing or not.

Moreover, in the case of $z=1$~(i.e., the standard Schwarzschild AdS black hole), when $\frac{S_y}{\mu}$ is small enough, the phase transition is always the second-order one until $\frac{S_y}{\mu}$ increases to a translating value, beyond which it switches to the first-order one, which is consistent with the results in Refs.~\cite{Herzog:2008he,Basu:2008st,Sonner:2010yx,Arean:2010zw,ybwu2014}. In the case of $z=\frac{6}{5}$ in Fig.~\ref{TraPz}(a), there are two translating points, which separate the first-order transition from the second-order one. With the increasing $z$, the dashed-line part standing for the first-order transition shrinks gradually until it disappears, such as the cases of $z=\frac{7}{5}$ and $\frac{3}{2}$ where the transition is always of the second-order regardless of the value of $\frac{S_y}{\mu^{1/z}}$, which means that the increasing $z$ hinders the emergence of the translating point. In fact, in the case of $\Delta_+=2$, we also observe the simultaneous appearance of the two translating points, such as $z=\frac{21}{20}$, which is similar to the case of $z=\frac{6}{5}$ in Fig.~\ref{TraPz}(a).

\begin{figure}
\begin{minipage}[!htb]{0.45\linewidth}
\centering
\includegraphics[width=2.9in]{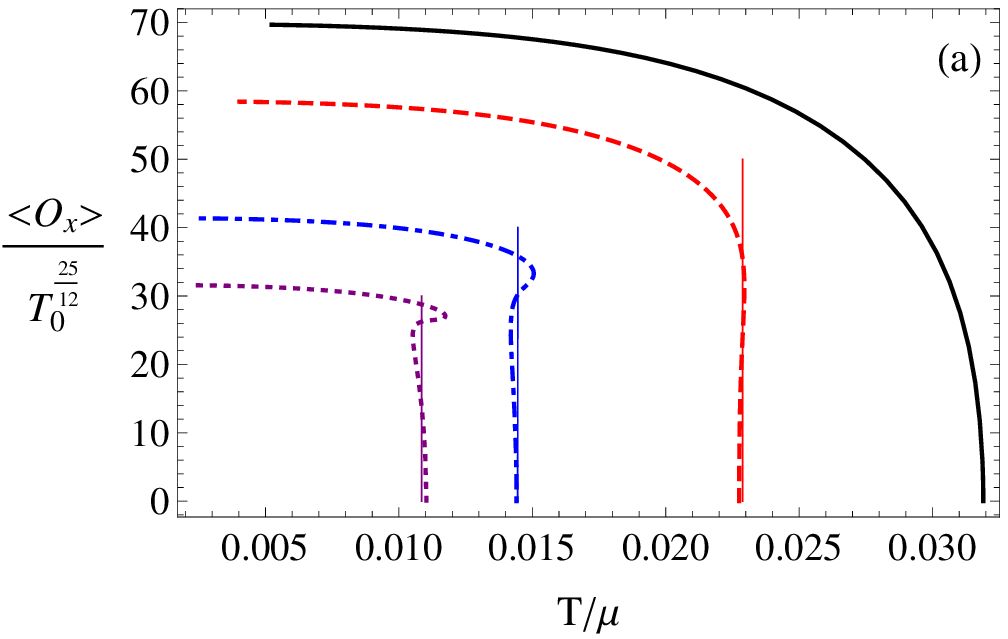}
\end{minipage}
\begin{minipage}[!htb]{0.45\linewidth}
\centering
\includegraphics[width=2.9in]{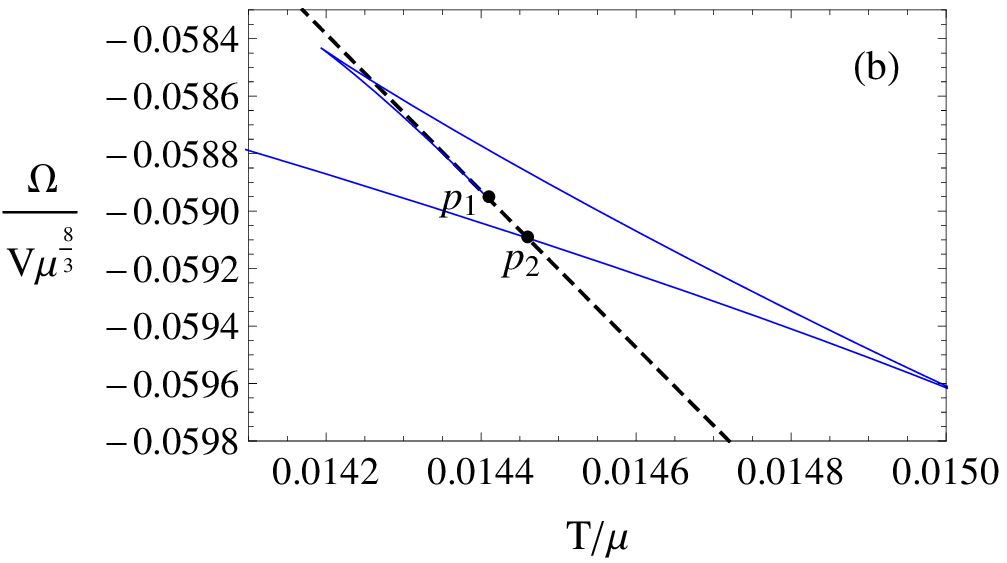}
\end{minipage}
\caption{The condensates (a)  and the grand potential (b) versus the temperature. The curves in (a) from top to bottom correspond to $\frac{S_y}{\mu^{5/6}}=\frac{1}{5}$,~$\frac{2}{5}$, $\frac{3}{5}$, and $\frac{7}{10}$, respectively, while the curves in (b) to the normal phase (dashed) and the superfluid phase with $\frac{S_y}{\mu^{5/6}}=\frac{3}{5}$  (solid), respectively, where $z=\frac{6}{5}$ and $\Delta_+=\frac{3}{2}$.}
\label{D3f2z6f5Con}
\end{figure}

Due to the rich phase structure, we display the condensates versus the temperature for the different values of  $\frac{S_y}{\mu^{5/6}}$ in the case of $z=\frac{6}{5}$ and $\Delta_+=\frac{3}{2}$ in Fig.~\ref{D3f2z6f5Con}.
 Just as we have predicted from Fig.~\ref{TraPz}, in the case of $\frac{S_y}{\mu^{5/6}}=\frac{1}{5}$~($\frac{2}{5}$), the phase transition is of the second~(first) order. In the case of $\frac{S_y}{\mu^{5/6}}=\frac{3}{5}$, we can observe that when the vector field begins to condensate, it seems a second-order phase transition as expected from the phase diagram. However, with the increasing condensate, the Cave of Winds observed only in the 5D AdS black hole with sufficiently large superfluid velocity~\cite{ybwu2014,Arean:2010zw} appears, which depends on the role of the Lifshitz parameter $z$ in changing the effective dimension of the gravitational background~\cite{Lu:2013tza}, and is obvious from the metric function, i.e., $f(r)$ in Eq.~(\ref{Lifm}). In this case, we have to calculate the grand potential to determine whether it is really the second-order transition. From Fig.~\ref{D3f2z6f5Con}(b), we see clearly that the system suffers originally a second-order transition at the point $p_1$. However, with the increasing condensate, a first-order phase transition takes place at the point $p_2$, hence, we conclude that the  phase transition is indeed of the first order. When the superfluid velocity improves gradually, we find the difference between the points $p_1$ and $p_2$ decreases and vanishes at the translating value $\frac{S_y}{\mu^{5/6}}\approx\frac{31}{50}$~(corresponding to the temperature $\frac{T}{\mu}\approx\frac{137}{10000}$), beyond which the phase transition is always of the second order, such as the case of $\frac{S_y}{\mu^{5/6}}=\frac{7}{10}$ in Fig.~\ref{D3f2z6f5Con}(a).

\begin{figure}
\begin{minipage}[!htb]{0.45\linewidth}
\centering
\includegraphics[width=2.9in]{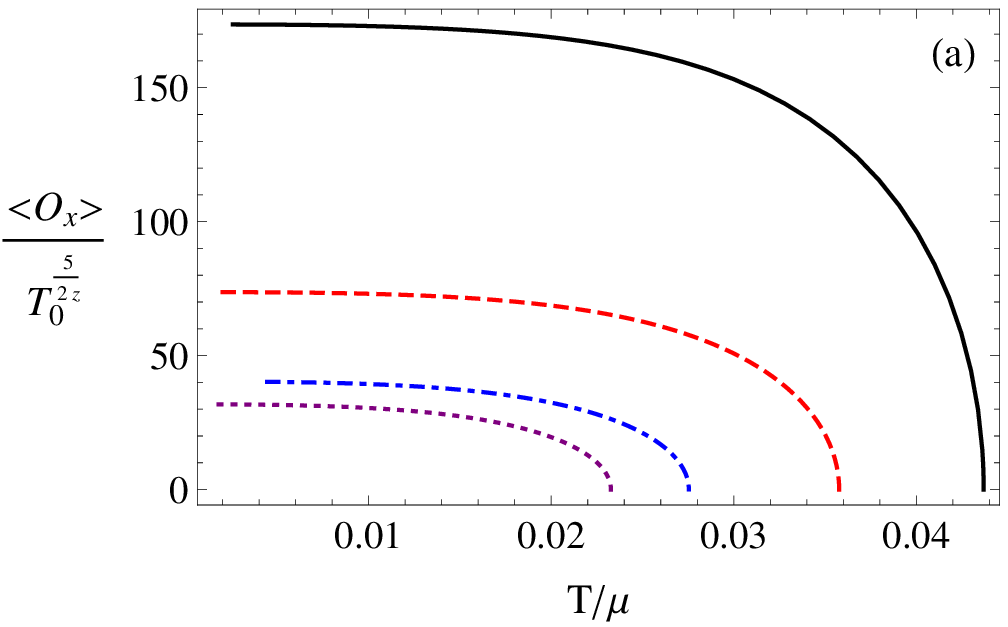}
\end{minipage}
\begin{minipage}[!htb]{0.45\linewidth}
\centering
\includegraphics[width=2.9in]{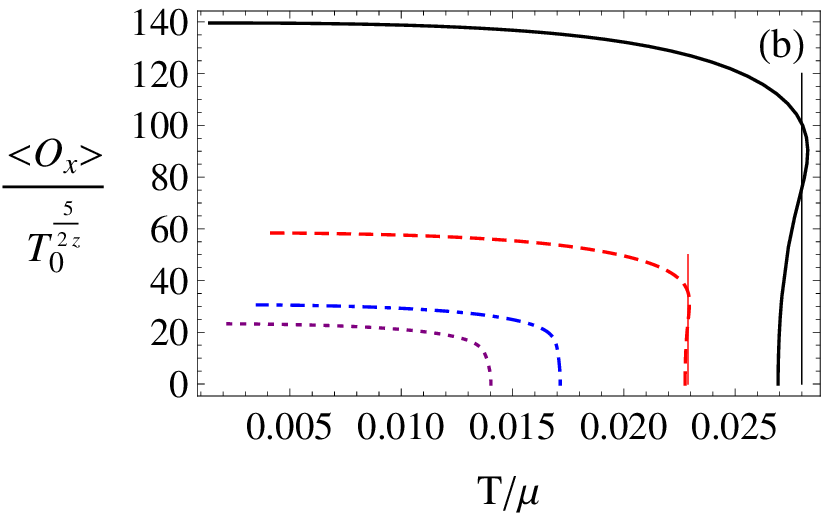}
\end{minipage}
\begin{minipage}[!htb]{0.45\linewidth}
\centering
 \includegraphics[width=2.9in]{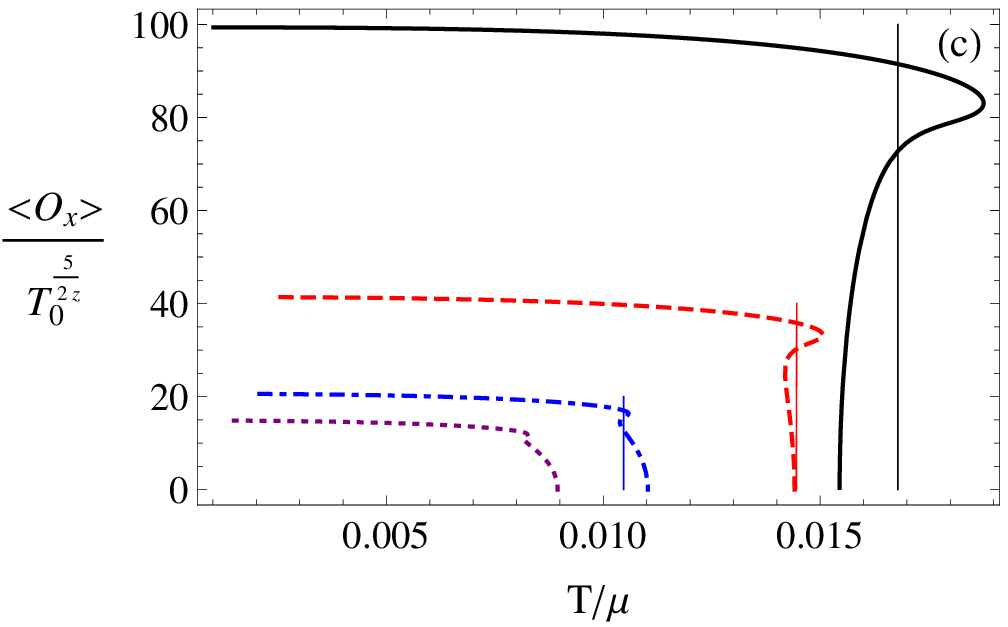}\\
\end{minipage}
\begin{minipage}[!htb]{0.45\linewidth}
\centering
\includegraphics[width=2.9in]{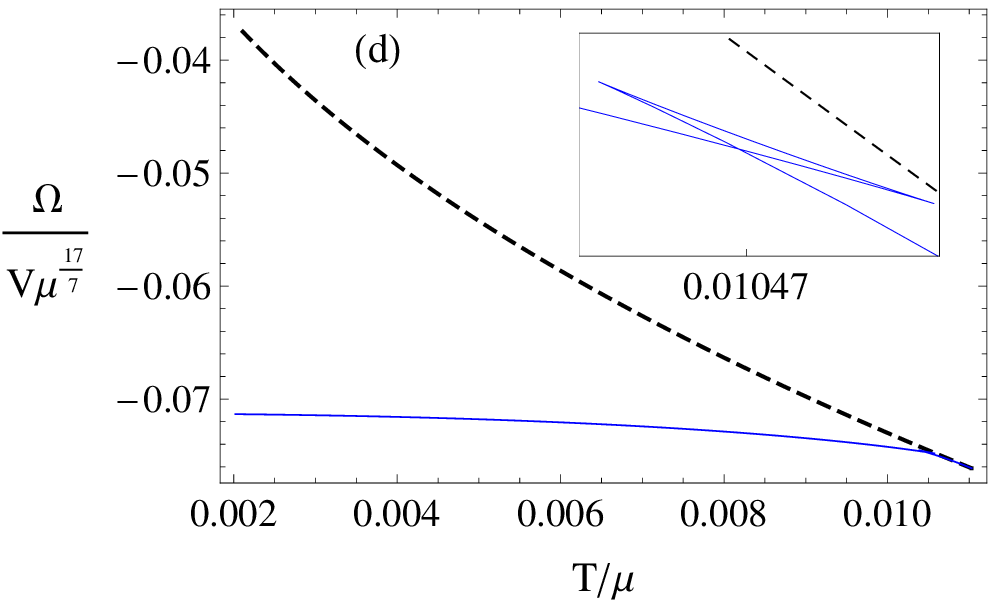}
\end{minipage}
\caption{The condensate versus the temperature for $\frac{S_y}{\mu^{1/z}}=0~$(a), $\frac{2}{5}$ (b), and $\frac{3}{5}$ (c). The curves in (a), (b) and (c) from top to bottom correspond to $z=1$, $z=\frac{6}{5}$, $z=\frac{7}{5}$, and $z=\frac{3}{2}$,  respectively.  The curves in (d)  stand for the grand potential versus the temperature for $z=\frac{7}{5}$ in the superfluid phase with $\frac{S_y}{\mu^{5/7}}=\frac{3}{5}$ (solid) and  the normal phase~(dashed), respectively.}
\label{D2Con}
\end{figure}

 Since the behaviors of the condensate with $\Delta_+=2$ are similar to the ones with $\Delta_+=\frac{3}{2}$, now we only plot the condensates versus the temperature for the different values of $z$ and $\frac{S_y}{\mu^{1/z}}$ with $\Delta_+=\frac{3}{2}$ in Fig.~\ref{D2Con}, from which we have the following remarks.
Fig.~\ref{D2Con}(a) is indeed the conductor/superconductor phase transition, from which we see clearly that the increasing $z$ can just hinder the phase transition and the condensate, and not change the order of the phase transition. What is more, at the lower temperature, the condensates saturate a fixed value decreasing with the improving $z$, which is consistent with the results in Refs.~\cite{Lu:2013tza,Wu:2014dta}. In the case of the lower superfluid velocity~($\frac{S_y}{\mu^{1/z}}=\frac{2}{5}$) in Fig.~\ref{D2Con}(b), the evident first-order transition emerges for $z=1$, which agrees with the result in Ref.~\cite{ybwu2014}. However, with the increasing $z$, the character of the first-order transition~(such as the double value of the condensate curve) becomes less obvious until the appearance of  the  second-order transition in the cases of $z=\frac{7}{5}$ and $\frac{3}{2}$, which not only means that the improving $z$ hinders the emergence of the first-order phase transition, but also agrees with the phase diagram in Fig.~\ref{TraPz}. For the larger superfluid velocity, for example, $\frac{S_y}{\mu^{1/z}}=\frac{3}{5}$ in Fig.~\ref{D2Con}(c), it is clear that the increasing $z$ makes the translation from the second-order transition to the first-order one more difficult. Moreover, due to the contribution of the Lifshitz parameter $z$ to the effective dimension of the gravitational spacetime, the Cave of Winds appears when taking $z=\frac{6}{5}$ and $\frac{7}{5}$, especially in the case of $z=\frac{6}{5}$ and $\frac{S_y}{\mu^{5/6}}=\frac{7}{10}$ in Fig.~\ref{D3f2z6f5Con}, which can only be observed in the 5D AdS black hole~\cite{ybwu2014,Arean:2010zw}.

 Because of the multiple-valued properties of the Cave of Winds,  the thermodynamically favored region should be determined via its grand potential. We typically plot the grand potential\footnote{It should be noted that the varying region of the grand potential for the Cave of Winds is too small to distinguish, hence, we here give a schematic contour with only a precisely stable bound.}  in the case of $z=\frac{7}{5}$ and $\frac{S_y}{\mu^{5/7}}=\frac{3}{5}$ in Fig.~\ref{D2Con}(d). It follows that  the grand potential in the superfluid phase is always less than the one in the normal phase, and the Cave of Winds has the similar intend to  the standard  one~\cite{ybwu2014,Arean:2010zw}. We have signed out the thermodynamically stable bound by a vertical line in Fig.~\ref{D2Con}(c), which has three intersecting points with the condensate curve. From the grand potential, we know
the region is unphysical between the points with larger and smaller values of condensate. Furthermore, we also sign out the stable bounds in the other multiple-valued condensates in Fig.~\ref{D3f2z6f5Con} and \ref{D2Con} by the same method. In addition, the fact that the Fig.~\ref{TraPz} is consistent with Figs.~\ref{D3f2z6f5Con} and \ref{D2Con} implies that the phase diagrams are rather reliable to reflect the critical behaviors.

\section{Supercurrents versus the superfluid velocity}
It is well known that Ginzburg-Landau~(GL) theory is the effective field theory of superconductors near the critical temperature~\cite{Hartnoll:2008kx}.  According to the free-energy density, it can give quite exact description and various significant quantities that can directly be compared with the experimental results. Especially, for the thin superconducting films or wires, the GL model indicates that, for example, the curve of supercurrent versus superfluid approximates a parabola opening downward; the squared ratio of the condensate with the critical current  to the condensate with vanishing superfluid velocity is equal to two thirds; the critical current is proportional to $(1-T/T_c)^{3/2}$ and so on. In addition, the gauge/gravity duality also indicates that our 4D gravity corresponds to the 3D field theory, i.e., the thin film in two spatial dimensions. Therefore, it is interesting to compare our results with the ones of the GL model to check the rationality of our holographic model.

\begin{figure}
\begin{minipage}[!htb]{0.45\linewidth}
\centering
\includegraphics[width=2.75in]{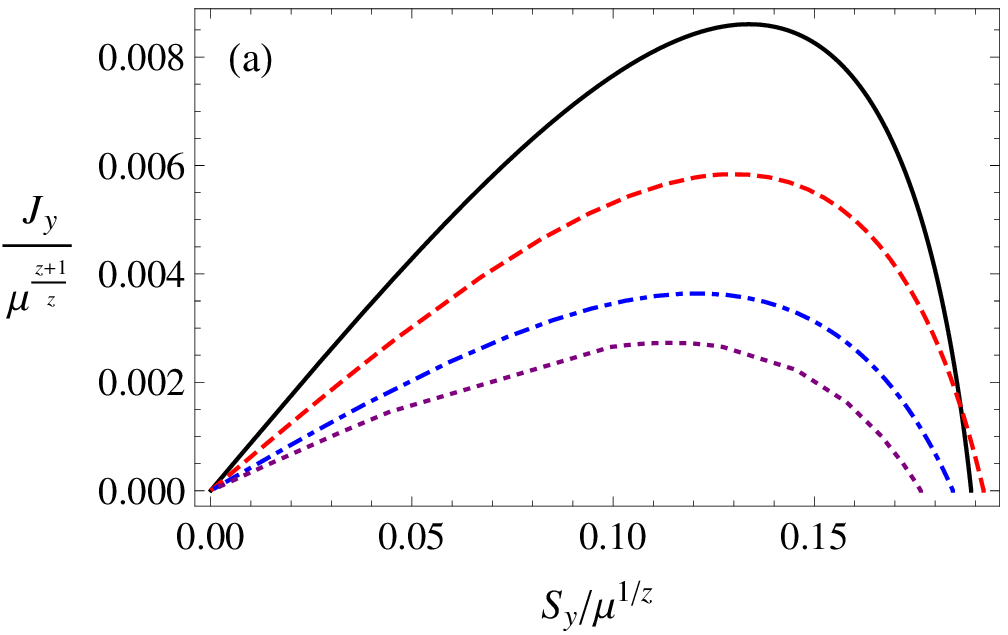}
\end{minipage}
\begin{minipage}[!htb]{0.45\linewidth}
\centering
 \includegraphics[width=2.75 in]{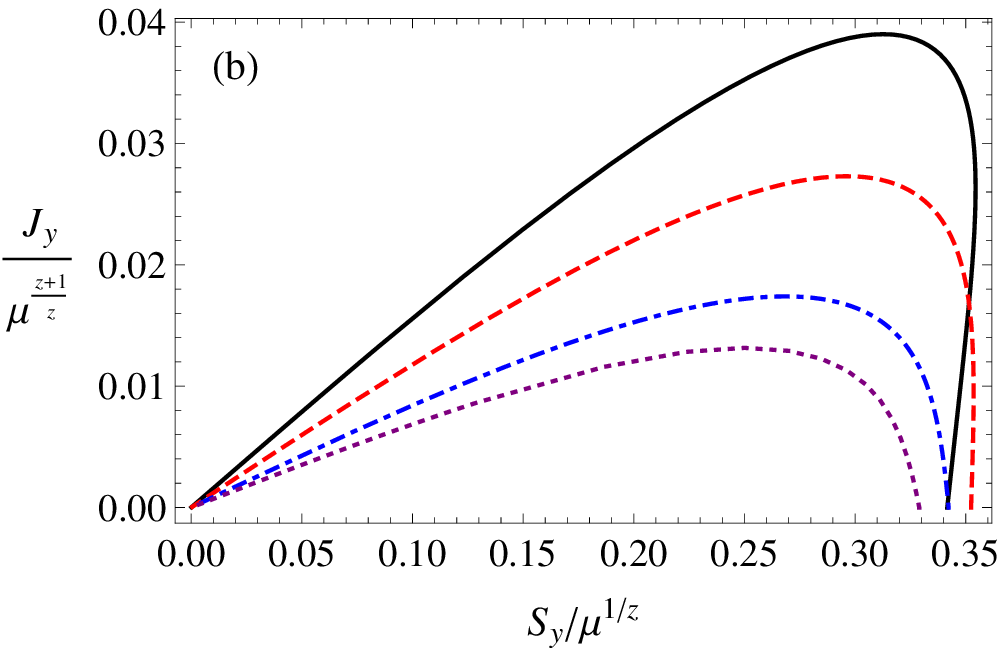}\\
\end{minipage}
\begin{minipage}[!htb]{0.45\linewidth}
\centering
\includegraphics[width=2.75in]{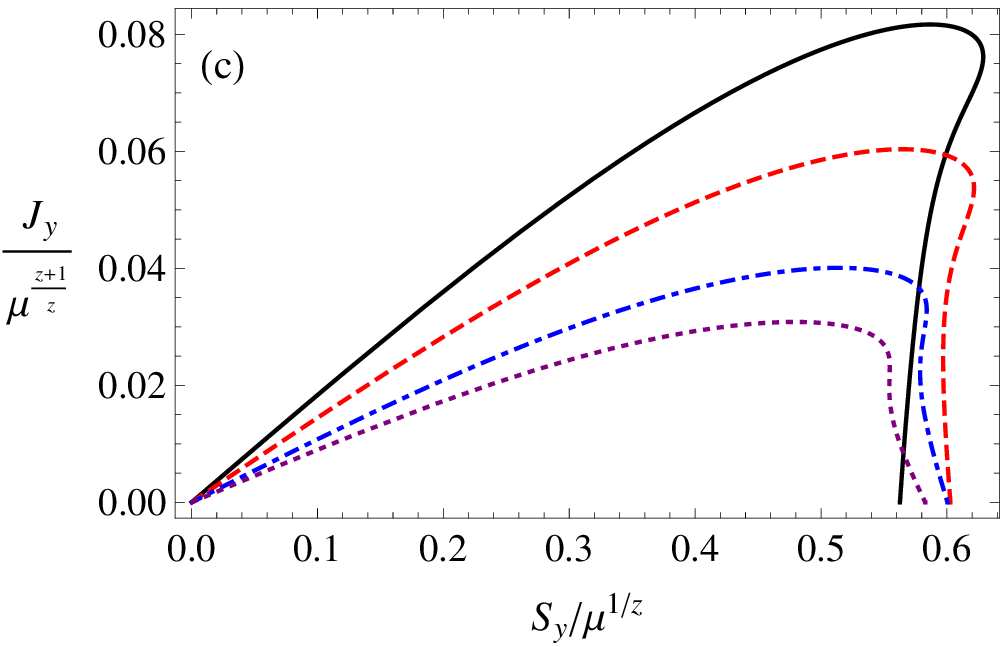}
\end{minipage}
\caption{The supercurrent versus the superfluid velocity for different values of the reduced temperature $\frac{T}{T_0}=\frac{9}{10}$~(a),~$\frac{7}{10}$~(b), and $\frac{2}{5}$~(c) with $\Delta_+=\frac{3}{2}$. The curves from top to bottom correspond to $z=1$, $\frac{6}{5}$, $\frac{7}{5}$, and $z=\frac{3}{2}$, respectively. }
\label{Del3f2CurSy}
\end{figure}

Because of the similar behaviors between $\Delta_+=\frac{3}{2}$ and $\Delta_+=2$, we typically plot the supercurrent as a function of the superfluid velocity with the different values of the reduced temperature and the Lifshitz parameter $z$ in the case of $\Delta_+=\frac{3}{2}$ in Fig.~\ref{Del3f2CurSy},
from which we have the following comments. Near the critical temperature~(i.e., $\frac{T}{T_0}=\frac{9}{10}$) in Fig.~\ref{Del3f2CurSy}(a),  the curves for all different $z$ approximate a parabola opening downward in accord with the GL model and have two intersecting points with the abscissa axis.  At the intersecting point with the larger superfluid velocity value denoted by $\frac{S_{yMax}}{\mu^{1/z}}$, the supercurrent  $J_y/\mu^{\frac{z+1}{z}}$ decreases smoothly to zero, which means that near the critical temperature, the system suffers a second-order phase transition agreeing  with the previous phase diagram and the condensate as well as the GL model. When the temperature falls over obviously from the critical temperature, such as $\frac{T}{T_0}=\frac{7}{10}$ in Fig.~\ref{Del3f2CurSy}(b), the linear dependence of $J_y/\mu^{\frac{z+1}{z}}$ on $\frac{S_{y}}{\mu^{1/z}}$ becomes more obvious until its maximum value $J_{yMax}/\mu^{\frac{z+1}{z}}$ (i.e., the critical current) comparing with the case of $\frac{T}{T_0}=\frac{9}{10}$, which is consistent with the Bardeen-Cooper-Schrieffer~(BCS) thin superconducting film~\cite{Tinkham1996}. Near $\frac{S_{yMax}}{\mu^{1/z}}$, the supercurrent for $z=1$ and $\frac{6}{5}$ becomes double valued, which means the latent heat and thus the first-order phase transition as expected from Fig.~\ref{TraPz}(a). As the temperature is lowered  sufficiently, such as $\frac{T}{T_0}=\frac{2}{5}$ in Fig.~\ref{Del3f2CurSy}(c), the interesting behavior still exists near $\frac{S_{yMax}}{\mu^{1/z}}$. We can easily see that near the critical point, the case of $z=1$ is the first-order transition, while the cases of $z=\frac{7}{5}$ and $\frac{3}{2}$ are of the second order. However, in the case of $z=\frac{6}{5}$~(i.e., the Cave of Winds), we can not determine intuitively the order of the transition.  By means of the phase diagram in Fig.~\ref{TraPz}(a), we conclude that the transition is of the first order, due to the fact that the critical superfluid velocity of $\frac{T}{T_0}=\frac{2}{5}$~(corresponding to $\frac{T}{\mu}\approx\frac{143}{10000}$)  is less than the translating value $\frac{S_y}{\mu^{5/6}}\approx\frac{31}{50}$ corresponding to the temperature  $\frac{T}{\mu}=\frac{137}{10000}$. Furthermore, for all cases in Fig.~\ref{Del3f2CurSy}(a), (b) and (c), the maximum value of the supercurrent decreases with the improving Lifshitz parameter $z$, which again implies that the larger $z$ hinders the phase transition.

Another well-known result of the GL model states that
\begin{equation}
\alpha:=\left(\frac{\langle O_x \rangle_c}{\langle O_x \rangle_\infty}\right)^2=\frac{2}{3},
\end{equation}
 where $\langle O_x \rangle_\infty$ and  $\langle O_x \rangle_c$ are denoted as the value of the condensate corresponding to the vanishing superfluid velocity and the critical current, respectively. Note that both $\langle O_x\rangle_c$ and $\langle O_x\rangle_\infty$ can be read from Fig.~\ref{Del3f2CurSy}. In particular, we can firstly fix the value of the critical current and thus obtain the corresponding condensate value, i.e., $\langle O_x\rangle_c$. Similarly, we can read off $\langle O_x\rangle_\infty$ from Fig.~\ref{Del3f2CurSy} with the superfluid velocity $S_y=0$.  We calculate the ratio $\alpha$ for different values of the reduced temperature and the Lifshitz parameter $z$ with $\Delta_+=\frac{3}{2}$, and list the results in Table~\ref{tab:GLpred},
\begin{table}
\caption{ The ratio $\alpha:=(\langle O_x\rangle_c/\langle O_x\rangle_\infty)^2$  at the different values of reduced temperature and the Lifshitz parameter with $\Delta_+=\frac{3}{2}$.}\label{tab:GLpred}
\begin{ruledtabular}
\begin{tabular}{c c c c c}
&$z=1$&$z=\frac{6}{5}$&$z=\frac{7}{5}$ & $z=\frac{3}{2}$ \\ \hline
$\frac{T}{T_0}=0.9$&$0.6619$&$0.6732$&$0.6427$ & $0.6480$ \\
$\frac{T}{T_0}=0.7$&$0.5832$&$0.5854$&$0.5829$ & $0.5850$ \\
$\frac{T}{T_0}=0.4$&$0.3011$&$0.3162$&$0.3328$ & $0.3399$ \\
\end{tabular}
\end{ruledtabular}
\end{table}
from which we find that, near the critical temperature~($\frac{T}{T_0}=\frac{9}{10}$), the ratio is consistent with the GL model, while in the case of the lower temperature~($\frac{T}{T_0}=\frac{7}{10},~\frac{2}{5}$), the ratio $\alpha$ deviates more evidently from the GL value $\frac{2}{3}$. For the fixed reduced temperature $\frac{T}{T_0}=\frac{2}{5}$, the ratio $\alpha$ increases with the improving $z$, which might result from the fact that the condensate $\langle O_x\rangle_\infty$ is suppressed more seriously as $z$ increases. However, in the cases of $\frac{T}{T_0}=\frac{7}{10}$ and $\frac{9}{10}$, the ratio $\alpha$ is not obviously dependent on the Lifshitz parameter $z$.
\section{Conclusion and discussion}
In summary, we have studied the holographic superfluid models in the 4D Lifshitz spacetime in the probe limit and obtained the effects of the Lifshitz dynamical exponent $z$  on the superfluid phase transition. Main results are concluded as follows.

We have obtained the phase diagrams with the rich structure and typically plotted the condensate as a function of the temperature. The results show that, regardless of the superfluid velocity, the critical temperature decreases with the improving Lifshitz  parameter $z$. Moreover, when $z$ improves beyond a value, there is no longer the first-order transition even in the case of the sufficiently large superfluid velocity, which is always observed in the usual AdS black hole in the probe limit. Hence, we conclude that the larger $z$ hinders not only the condensate but also the appearance of translating point from the second-order transition to the first-order one. What is more, in some range of the Lifshitz parameter $z$, the Cave of Winds appears, which has been observed in the 5D AdS black hole with the intermediate mass squared for some superfluid velocity. This means that the Lifshitz parameter $z$ contributes evidently to the effective mass as well as the effective dimension of the background geometry, which is obvious from the characteristic exponent $\Delta_\pm=\frac{1}{2}\left(z\pm\sqrt{z^2+4m^2}\right)$ and the spacetime metric (\ref{Lifm}).

Furthermore, we have compared our holographic model with the GL model by plotting the supercurrent versus the  superfluid velocity. The results from the supercurrent are consistent with the ones obtained from the condensate with the fixed superfluid velocity, especially the Cave of Winds.  In addition, we have shown that our results agree with the ones of the GL model near the critical temperature, especially, the ratio $\alpha=\left(\langle O_x \rangle_c/{\langle O_x \rangle_\infty}\right)^2$, but this value deviates much more obviously as the temperature decreases further.

Notice that our study of the MCV superfluid model was limited to the probe limit in the 4D spacetime and the effects of the Lifshitz parameter $z$ were discussed by comparing with the results in  the 5D AdS black hole~\cite{ybwu2014}.   To complement our results, it is interesting to consider the holographic superfluid model in the 5D Lifshitz spacetime as well as the backreaction from the MCV field. In addition, by applying the Landau criterion to the quasinormal mode identified with the pole of Green function, the authors of Ref.~\cite{Amado:2013aea} revisited the holographic superfluid, where the results exhibited the much lower critical temperature than the one from the thermodynamical analysis and also the signal of the existence for the striped phase near the critical point. Therefore, it is valuable to analyze the MCV model from the perturbing perspective to further understand the Lifshitz effect, which is our work in the near future.

\acknowledgments  We would like to take this opportunity to thank  L.~Li for his helpful discussions and comments. This work is supported by the National Natural Science Foundation of China (Grant No.~11175077), the Joint Specialized Research Fund for the Doctoral Program of Higher Education, Ministry of Education, China (Grant No.~20122136110002), the Project of Key Discipline of Theoretical Physics of Department of Education in Liaoning Province (Grant Nos.~905035 and 905061), and the Open Project Program of State Key Laboratory of Theoretical Physics, Institute of Theoretical Physics, Chinese Academy of Sciences, China (No.~Y4KF101CJ1).


\begin{thebibliography}{*}
\bibitem{Maldacena1998}J.~M.~Maldacena,  Adv. Theor. Math. Phys. 2 (1998) 231.
\bibitem{Hartnoll2008}S.~A.~Hartnoll, C.~P.~Herzog and G.~T.~Horowitz,  Phys. Rev. Lett. 101 (2008) 031601.
\bibitem{Gubser2008a} S.~S.~Gubser and S.~S.~Pufu, JHEP 11 (2008) 033.
\bibitem{Chen:2010mk} J.~W.~Chen, Y.~J.~Kao, D.~Maity, W.~Y.~Wen and C.~P.~Yeh,
     Phys.\ Rev.\ D 81 (2010)  106008.
\bibitem{Hartnoll:2008kx}  S.~A.~Hartnoll, C.~P.~Herzog and G.~T.~Horowitz, JHEP 0812 (2008) 015.
\bibitem{Cai:2014oca}  R.~-G.~Cai and R.~-Q.~Yang,  arXiv:1404.2856 [hep-th].
\bibitem{Zeng:2014uoa}  H.~B.~Zeng and J.~P.~Wu,  Phys.\ Rev.\ D  90 (2014) 046001.
\bibitem{Cai:2014jta}  R.~-G.~Cai and R.~-Q.~Yang,    arXiv:1404.7737 [hep-th].
\bibitem{Albash:2008eh}  T.~Albash and C.~V.~Johnson, JHEP  0809 (2008) 121.
\bibitem{Liu:2011fy} Y.~Liu, Q.~Pan and B.~Wang,  Phys.\ Lett.\ B 702 (2011) 94.
\bibitem{Momeni:2012ab}  D.~Momeni, N.~Majd and R.~Myrzakulov, Europhys.\ Lett.\ 97 (2012) 61001.
\bibitem{Nishioka131}T.~Nishioka, S.~Ryu and T.~Takayanagi,   JHEP 03 (2010) 131.
\bibitem{Roychowdhury:2012vj}  D.~Roychowdhury,  Phys.\ Lett.\ B  718 (2013) 1089.
 \bibitem{Herzog:2009xv}  C.~P.~Herzog,  J.\ Phys.\ A  42 (2009) 343001.
\bibitem{Horowitz:2010gk}  G.~T.~Horowitz,  Lect.\ Notes Phys.\  828 (2011) 313.
 \bibitem{Kachru}S.~Kachru, X.~Liu and M.~Mulligan,   Phys.\ Rev.\ D 78 (2008) 106005.
\bibitem{Horava:2009uw}  P.~Horava,  Phys.\ Rev.\ D 79 (2009) 084008.
 \bibitem{Liu:2013kla}  M.~Liu, J.~Lu, Y.~Xu, J.~Lu, Y.~Wu and R.~Wang,  Phys.\ Rev.\ D 87 (2013) 024043.
\bibitem{Liu:2014lga}  M.~Liu, Y.~Xu, J.~Lu, Y.~Yang, J.~Lu and Y.~Wu,   Mod.\ Phys.\ Lett.\ A  29 (2014) 1450084.
\bibitem{Iizuka:2012wt}  N.~Iizuka and K.~Maeda,   JHEP 1207 (2012) 129.
\bibitem{Koga:2014hwa}  J.~i.~Koga, K.~Maeda and K.~Tomoda,   Phys.\ Rev.\ D 89 (2014) 104024.
\bibitem{Pdw09052678}   D.~W.~Pang,   Commun.\ Theor.\ Phys.\  62 (2014) 265.
\bibitem{Brynjolfsson065401} E.~J.~Brynjolfsson, U.~H.~Danielsson, L.~Thorlacius and T.~Zingg,   J.\ Phys.\ A 43 (2010) 065401.
\bibitem{Sin4617}S.~-J.~Sin, S.~-S.~Xu and Y.~Zhou, Int.\ J.\ Mod.\ Phys.\ A 26 (2011) 4617.
 \bibitem{Buyanyan}Y.~Bu,  Phys.\ Rev.\ D 86 (2012) 046007.
\bibitem{Lu:2013tza}  J.~W.~Lu, Y.~B.~Wu, P.~Qian, Y.~Y.~Zhao, X.~Zhang and N.~Zhang,  Nucl.\ Phys.\ B  887 (2014) 112.
\bibitem{Zhao:2013pva}   Z.~Zhao, Q.~Pan and J.~Jing,  Phys.\ Lett.\ B 735 (2014) 438.
\bibitem{Momeni:2012tw}   D.~Momeni, R.~Myrzakulov et al,  Int. J. Geom. Methods Mod. Phys. 12 (2015)1550015.
\bibitem{Lala:2014jca} A.~Lala,  Phys.\ Lett.\ B 735 (2014) 396.
\bibitem{Tallarita:2014bga}  G.~Tallarita,   Phys.\ Rev.\ D 89 (2014) 106005.
\bibitem{Lin:2014bya}  K.~Lin, E.~Abdalla and A.~Wang,  arXiv:1406.4721 [hep-th].
\bibitem{Guo:2014wca}  H.~Guo, F.~W.~Shu, J.~H.~Chen, H.~Li and Z.~Yu, arXiv:1410.7020 [hep-th].
\bibitem{Jing:2014bza}  J.~Jing, S.~Chen and Q.~Pan,  arXiv:1410.7847 [hep-th].
\bibitem{Herzog:2008he}  C.~P.~Herzog, P.~K.~Kovtun and D.~T.~Son,   Phys.\ Rev.\ D 79 (2009) 066002.
\bibitem{Basu:2008st}  P.~Basu, A.~Mukherjee and H.~-H.~Shieh,  Phys.\ Rev.\ D 79 (2009) 045010.
\bibitem{Amado:2009ts}  I.~Amado, M.~Kaminski and K.~Landsteiner,   JHEP 0905 (2009) 021.
 \bibitem{Sonner:2010yx}   J.~Sonner and B.~Withers, Phys.\ Rev.\ D 82 (2010) 026001.
\bibitem{Arean:2010xd}  D.~Arean, M.~Bertolini, J.~Evslin and T.~Prochazka,  JHEP 1007 (2010) 060.
\bibitem{Zeng:2010fs}  H.~-B.~Zeng, W.~-M.~Sun and H.~-S.~Zong,  Phys.\ Rev.\ D 83 (2011) 046010.
\bibitem{Arean:2010zw}  D.~Arean, P.~Basu and C.~Krishnan,   JHEP  1010 (2010) 006.
\bibitem{Amado:2013xya}  I.~Amado, D.~Arean, A.~Jimenez-Alba, K.~Landsteiner, L.~Melgar and I.~S.~Landea,  JHEP  1307 (2013) 108.
\bibitem{Amado:2013aea}  I.~Amado, D.~Are¨¢n, A.~Jim¨¦nez-Alba, K.~Landsteiner, L.~Melgar and I.~Salazar Landea,  JHEP 1402 (2014) 063.
\bibitem{Cai:2013pda} R.~G.~Cai, S.~He, L.~Li and L.~F.~Li,   JHEP 1312 (2013) 036.
\bibitem{Cai:2013kaa}  R.~-G.~Cai, L.~Li, L.~-F.~Li and Y.~Wu,  JHEP  1401 (2014) 045.
 \bibitem{Wu:2014dta}  Y.~B.~Wu, J.~W.~Lu, M.~L.~Liu, J.~B.~Lu, C.~Y.~Zhang and Z.~Q.~Yang,   Phys.\ Rev.\ D 89 (2014) 106006.
\bibitem{Wu:2014lta} Y.~B.~Wu, J.~W.~Lu, Y.~Y.~Jin, J.~B.~Lu, X.~Zhang, S.~Y.~Wu and C.~Wang,  Int.\ J.\ Mod.\ Phys.\ A 29 (2014) 1450094.
 \bibitem{Cai:2013aca}  R.~-G.~Cai, L.~Li and L.~-F.~Li,   JHEP 1401 (2014) 032.
\bibitem{Li:2013rhw}  L.~-F.~Li, R.~-G.~Cai, L.~Li and C.~Shen,  arXiv:1310.6239 [hep-th].
\bibitem{Cai:2014ija}  R.~-G.~Cai, L.~Li, L.~-F.~Li and R.~-Q.~Yang,   JHEP  1404 (2014) 016.
 \bibitem{ybwu2014} Y.~B.~Wu, J.~W.~Lu, W.~X.~Zhang, C.~Y.~Zhang, J.~B.~Lu and F.~Yu,  arXiv:1410.5243 [hep-th].
\bibitem{Chernodub:2011mc}  M.~N.~Chernodub,   Phys.\ Rev.\ Lett.\  106 (2011) 142003.
\bibitem{Bali:2011qj}  G.~S.~Bali, F.~Bruckmann, G.~Endrodi, Z.~Fodor, S.~D.~Katz, S.~Krieg, A.~Schafer and K.~K.~Szabo,
  JHEP 1202 (2012) 044.
\bibitem{Tinkham1996} M.~Tinkham,  Introuduction to Superconducivity (McGraw-Hill, New York, 1996).
\end{thebibliography}
\end{document}